\documentclass{article}
\usepackage{epsfig}

\begin{document}

\title{Vortex induced confinement and the IR properties of
Green functions\footnote{\uppercase{T}alk presented by 
\uppercase{K}.~\uppercase{L}angfeld at ``Confinement V'', Gargano,
Italy, 10-14 Sept 2002.  }}

\author{K.~Langfeld, J.C.R.~Bloch, J.~Gattnar, H.~Reinhardt \\ \\ 
Institut f\"ur Theoretische Physik, \\ 
Universit\"at T\"ubingen, D-72076 T\"ubingen, Germany \\ \\
A.~Cucchieri and  T.~Mendes\footnote{\uppercase{W}ork 
supported by \uppercase{FAPESP}, \uppercase{B}razil 
(\uppercase{P}roject \uppercase{N}o.\ 00/05047-5).} \\ \\ 
IFSC S\~ao Paulo University \\
 C.P. 369 CEP 13560-970, S\~ao Carlos (SP), Brazil }


\maketitle

In order to observe the confinement of quarks and gluons in 
Yang-Mills theory, two conditions must necessarily be 
obeyed~\cite{wj2000}: 
(i) there must be a long range order of certain un--physical 
degrees of freedom ({\it violation of the cluster decomposition }); 
(ii) there must be a mass gap for physical excitations.
Using Landau gauge where the gauge 
configurations are restricted to the first Gribov regime, 
it was firstly speculated by Gribov~\cite{Gribov:1977wm} that the divergence 
of the ghost propagator at zero momentum transfer signals
confinement. Over the years, Zwanziger has put forward that the 
configurations which are relevant for confinement are close 
to the Gribov horizon\cite{Zwanziger:1991ac}. A formal relation
between confinement and the IR properties of confinement was 
advocated by Kugo and Ojima~\cite{Kugo:gm}: their approach is 
based on the assumption that a unique definition of the BRST charge 
operator exists. This assumption, however, is invalidated by the 
presence of Gribov copies. 

\vskip 0.2cm
In my talk, I will explore the relation between the IR properties 
of Green functions and confinement using SU(2) lattice gauge theory. 

\vskip 0.2cm
\begin{figure}[t]
%\epsfxsize=10cm   %width of figure - will enlarge/reduce the figures
%\epsfbox{fig3.eps}
%\figurebox{2cm}{3cm}{} %to have a box alone 
\centerline{
\epsfxsize=5.2cm\epsfbox{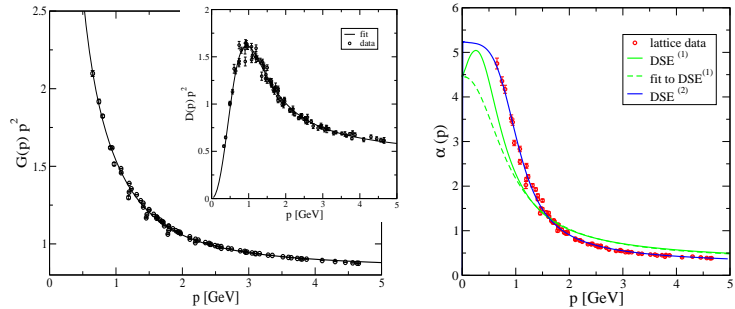}
\hspace{0.2cm}
\epsfxsize=4cm\epsfbox{alpha_all.eps}}   
\caption{ Ghost-- and gluon--form factor (left), running coupling
(right).  \label{fig:1}}
\end{figure}

Recent investigations of the Dyson-Schwinger 
equations~\cite{vonSmekal:1997is,Atkinson:1997tu} 
have revealed that the ghost--propagator $G(p^2)$ and the 
gluon--propagator $D(p^2)$ satisfy remarkable scaling relations 
in the IR limit: 
\begin{equation}
G(p^2) \sim p^{-2 \, - \, 2 \kappa } \; , \; \; \; 
D(p^2) \sim p^{-2 \, + \, 4 \kappa } \; ,  
\label{eq:1}
\end{equation}
where $\kappa \in [0.5, 1] $ depending on the truncations. 
Some of our lattice results for these propagators are shown in
figure~\ref{fig:1} (see also~\cite{Bloch:2002we}). 
The lattice data are consistent with an 
IR diverging form factor with $\kappa \approx 0.5$. 
This finding is in agreement with a large scale volume study 
of the Adelaide group~\cite{Bonnet:2001uh}. The IR scaling relation 
(\ref{eq:1}) implies that the running coupling 
$ 
\alpha (p^2) \; = \; \alpha \; \Bigr[ p^2 D(p^2) \Bigl] \; 
\Bigr[ p^2 G(p^2) \Bigl]^2 
$ 
approaches a constant for $p \rightarrow 0$. In the present lattice 
simulations, we have not yet observed the flattening of the coupling 
at low momenta. Additional theory input is required to detect 
the fixed point value $\alpha (0)$. Two truncations of the DSE 
system are presently available: DSE$^{(1)}$~\cite{Fischer:2002hn}, 
and a truncation scheme using a 2-loop inspired kernel and using 
$\kappa =0.5$ (DSE$^{(2)}$)~\cite{bloch}. Using the latter DSE solution 
as an additional theory input, our lattice data are consistent with the fixed 
point value: $\alpha (0) \; = \; 5.2 \pm 0.3$. 

\begin{figure}[t]
%\epsfxsize=10cm   %width of figure - will enlarge/reduce the figures
%\epsfbox{fig3.eps}
%\figurebox{2cm}{3cm}{} %to have a box alone 
\centerline{
\epsfxsize=5cm\epsfbox{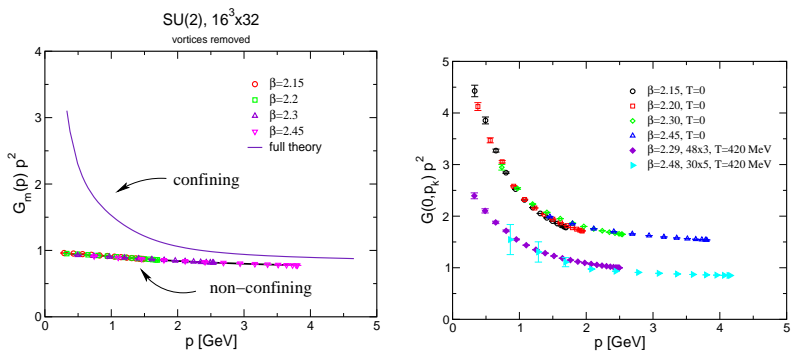}
\hspace{0.2cm}
\epsfxsize=5cm\epsfbox{ghost_t.eps}}   
\caption{ Ghost--form factor of the non-confining theory (left), 
   Ghost--form factor at finite temperatures (right).  \label{fig:2}}
\end{figure}

\vskip 0.2cm
It was recently observed~\cite{Langfeld:2002bg} that removing the 
percolating vortex cluster from the lattice configurations by hand 
implies quark de-confinement. At the same time, the 
IR divergence of the ghost form factor is removed (see figure 
\ref{fig:2}). Here, we present for the first time the ghost
form-factor $G(0, \vec{p})$ at temperatures well above the 
de-confinement phase transition (see figure \ref{fig:2}). 
Using dimensional reduction, we know that the 3-dimensional 
reduced theory describing Yang-Mills theory at asymptotic
temperatures, is in the confining phase. Hence, we expect that the 
ghost form-factor is still diverging for $\vec{p}^2 \rightarrow 0$. 
Our lattice result meets with this expectation. 

\vskip 0.2cm
We argue that the removal of the percolating cluster of confining 
vortices restores cluster decomposition: the lattice 
study~\cite{Langfeld:2002bg} shows that, in this case, the 
theory is converted in a non-confining theory whose ghost form-factor 
is regular in the IR limit. 
It was observed in the recent past~\cite{Langfeld:1998cz} 
that the de-confinement phase
transition at finite temperatures can be viewed as vortex 
de-percolation transition: considering a hypercube spanned by 
the time and 2-spatial axis, the vortex world lines stop percolating 
at $T_c$. Cluster de-composition is partially restored. 
At the same time, the spatial hypercube is still filled by 
percolating vortex world lines. As a result, the violation of cluster 
decomposition of the 3-dimensional reduced theory by means of percolating 
vortex world lines is still active (as requested for an explanation 
of the spatial string tension).

\end{document}